\documentclass[epj]{webofc}
\usepackage[varg]{txfonts,epsfig}   
\wocname{EPJ Web of Conferences}
\woctitle{INPC 2013}
\begin{document}
\title{On the conversion of neutron stars into quark stars}
%
%

\author{Giuseppe Pagliara\inst{1,2}\fnsep\thanks{\email{pagliara@fe.infn.it}} 
}

\institute{Dipartimento di Fisica e Scienze della Terra dell'Universit\`a di Ferrara, Via Saragat 1, I-44100 Ferrara, Italy
\and
           INFN
Sez.~di
Ferrara, Via Saragat 1, I-44100 Ferrara, Italy}

\abstract{The possible existence of two families of compact stars, neutron stars and quark stars,
naturally leads to a scenario in which a conversion process between the two stellar objects occurs
with a consequent release of energy of the order of $10^{53}$ erg. We discuss recent
hydrodynamical simulations of the burning process and neutrino diffusion simulations of cooling of a
newly formed strange star. We also briefly discuss this scenario in connection with recent 
measurements of masses and radii of compact stars.}
\maketitle
\section{Introduction}
\label{intro}
The equation of state and the properties of nuclear matter at high baryon
densities ($n_B > 3 n_0$ with the nuclear saturation density
$n_0=0.16$ fm$^{-3}$ ) are rather uncertain due to the lack of
laboratory experimental data and to the theoretical and computational
difficulties in the study of such a quantum many body relativistic
problem. On the other hand at high temperature and vanishingly small
baryon density a consistent physical picture has emerged in the
last years thanks to experiments of heavy ion collisions and to the
complicated calculations of lattice QCD. In particular what is now
widely accepted is the existence of a cross-over transition from
hadronic matter to the quark gluon plasma (a soup of quarks and
gluons) at a critical temperature $T_c \sim 170$ MeV.  While the
matter created in the early stages of ultra-relativistic heavy ion
collisions is deconfined, the subsequent expansion and cooling down
causes hadronization; the production of particles, at the so called
freeze-out, is extremely well described by hadron resonance gas models
which consist of a mixture of baryons and mesons, with masses up to $\sim 2$
GeV, in thermodynamical equilibrium. In particular, in the baryon
sector the contribution of hyperons is important and cannot be
neglected. It is interesting to notice that the equation of state
associated with a hadron resonance gas has an almost constant sound
velocity of $\sim \sqrt{1/6}$ and it is therefore {\it softer} than the
equation of state of the quark gluon plasma which, at least in the
limit of very high temperatures, can be modelled by a gas of massless
quarks and gluons with a corresponding sound velocity of $\sim \sqrt{1/3}$ 
(see \cite{Kolb:2003dz} for a review).

While, as explained before, at high baryon densities there are huge
uncertainties on the equation of state, from the high temperature
regime on can learn two important lessons: i) at high energy densities
a phase transition to quarks and gluons matter occurs ii) at smaller
energy densities not only pions and nucleons are present but also
hyperons, resonances and strange mesons. Moreover the equation of
state of the hadronic phase is softer than the one of the quark
phase due to the appearance of new degrees of freedom
when increasing the temperature. A natural place in which nuclear matter at high density exists
is the core of neutron stars. Recent studies and observations of these
stellar objects have severely challenged nuclear physics. We refer in
particular to the very precise measurements of neutron stars with
masses up to two solar masses \cite{Demorest:2010bx}. Moreover,
although the corresponding measurements are still affected by large
uncertainties, there are several indications of stars having a radius
smaller than about $10$ km \cite{Guillot:2013wu,Lattimer:2013hma}. It
is not easy to explain the existence of massive neutron stars and very
compact stars with only one family of compact objects, indeed a star
with a mass of $2 M_{\odot}$ and a radius of $9$ km would be close to
the causal limit (i.e. sound velocity equal to $c$) and thus unlikely
\cite{Lattimer:2006xb}. It is possible instead that two families
actually exist differing from each other for the composition: hadronic
stars made of baryonic matter and stars made of quark matter
\cite{Berezhiani:2002ks,Bombaci:2004mt,Drago:2013fsa}. If these two families exist,
there must be a mechanism for populating the quark star branch of
compact stars. We do not investigate here in detail the astrophysical
scenario of the conversion of a neutron star into a quark star but
just assume that it can take place and discuss instead if such a
process can be phenomenologically relevant.

\section{Hydrodynamical conversion and neutrino emission}
The conversion of a neutron star into a quark star has been
investigated in many papers (see \cite{Drago:2005yj,Niebergal:2010ds} and references
therein) and it has been shown to proceed as a strong deflagration
i.e. it is similar to a burning process which is sustained by the
energy released in the conversion and does not propagate as a shock
wave. Interestingly, the hydrodynamical instabilities which develop,
mainly because of the gradient of the gravitational potential inside the star,
significantly increase the rate of conversion. In
\cite{Herzog:2011sn,Pagliara:2013tza} full 3+1D hydrodynamical
simulations of the burning process have been performed in which the
conversion is described in a discontinuity approximation, meaning that
the unburned material (nucleonic matter) and the newly converted
strange quark matter are separated by a conversion front. The
conversion process is triggered by a strange quark matter seed in the
center of the initial stellar configuration: a volume in the center is
converted instantly and the conversion front subsequently propagates
as a deflagration wave. The development of a turbulent flow strongly 
enhances the velocity of conversion.

\begin{figure}[ptb]
\vskip 0.5cm
\begin{centering}
\epsfig{file=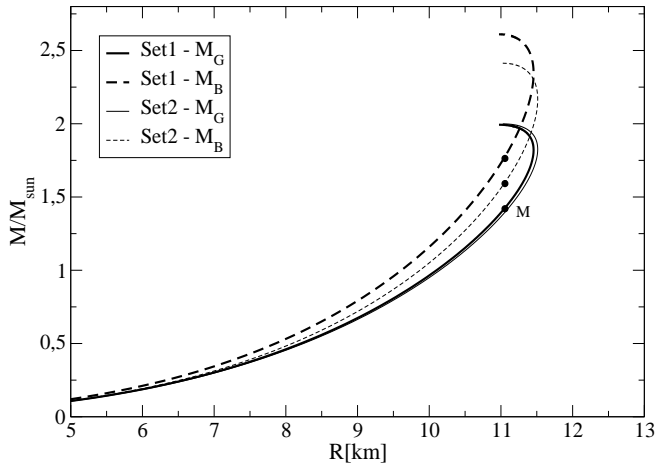,height=8.5cm,width=6cm,angle=-90}
\caption{Gravitational mass and baryonic mass of quark stars obtained
  with two parametrized equations of state.  The larger value of
  quarks binding energy in bulk matter at zero pressure obtained with
  Set1 provides, correspondingly, a larger value of the total
  binding energy of quark stars.}
\end{centering}
\end{figure}

The neutron star is modeled by use of the Lattimer-Swesty 
equation of state with compressibility $K=240$ MeV \cite{Lattimer:1991nc}
and for quark stars, as customary, we adopt the simple 
MIT bag model picture. We consider in particular two sets of
parameters, taken from \cite{Weissenborn:2011qu}, both allowing to reach
a maximum mass for strange quark stars of $2 M_{\odot}$. We
fix the current mass of the strange quark to $m_s = 100$
MeV, and we then consider $B^{1/4}_{eff} = 142$ MeV - $a_4$ = 0.9
(Set1) and $B^{1/4}_{eff} = 141$ MeV - $a_4$ = 0.65 (Set2) where $B_{eff}$
and $a_4$ represent the effective bag constant and the coefficient of the 
$\mu^4$ term in the pressure of the quark phase,
respectively ($\mu$ is the quark chemical potential), as in
Ref. \cite{Alford:2004pf}. The two sets provide almost the 
same mass radius relation, as one can notice in Fig.1, but they correspond to different values of the
energy per baryon of the ground state of strange quark
matter: for Set1 $E/A$ = 860 MeV and for Set2 $E/A$ = 930
MeV implying that the former provides quark bags with a larger binding energy. 
The larger binding energy 
of quarks in bulk matter obtained with Set1 
is then responsible for the larger total binding 
energy (the difference between the baryonic mass $M_B$ and the gravitational mass $M_G$ \cite{Bombaci:2000cv}) 
of quark stars: for instance the stellar
configuration labelled with M, which lies on the two gravitational mass
radius curves, has a larger baryonic mass (the point on the thick dashed
line) in the case of Set1. In turn, this means that a larger energy is available for the
conversion of a neutron star when the equation of state of Set1 is
adopted. As previously discussed, the burning proceeds thanks to the
energy released in the conversion and indeed while for Set1 a
successful conversion is obtained, for Set2 the energy released in the
conversion is not sufficient to power the conversion front and the
burning stops immediately after the initialisation.  The conversion,
when successful, is very efficient: turbulence indeed leads to very
high conversion velocities and on a time scale of few milliseconds
almost the whole star is converted.

The newly born strange quark star is hot due to the huge energy
released during the conversion process. The temperature at the center
of the star reaches values up to $40$ MeV and decreases smoothly as a
function of the radius until the layer separating the burned from
the unburned material. In correspondence of this interface a steep
temperature gradient is obtained (see Ref.\cite{Pagliara:2013tza})
which, within the diffusion approximation, leads to a fast neutrino cooling.

For an order of magnitude estimate let us consider the inverse mean
free path for the scattering of neutrinos off degenerate quarks \cite{Steiner:2001rp}
$\frac{1}{\lambda_{s}}=\frac{G_F^2E_\nu^3\mu_i^2}{5\pi^3}$ where
$G_F=1.17\times 10^{-5}$ GeV$^{-2}$ is the Fermi constant, $E_\nu$ is
the energy of the neutrino/antineutrino and $\mu_i$ is the chemical
potential of the particle involved in the scattering (up, down,
strange quarks). For mean thermal energy neutrinos, $E_\nu=\pi T$, with $T
\sim 50$ MeV and $\mu_i \sim 500$ MeV at the center of the star, thus
one can estimate the diffusion time as $\tau \sim R^2/\lambda_s$ where $R
\sim 10$ km is the radius of the star. We obtain a time scale of the
order of $1$s. Actually in the diffusion calculations presented in
\cite{Pagliara:2013tza} the star cools down to temperatures below $1$
MeV after about $10-15$ s with a initial neutrino luminosity of
$10^{52}$ erg/s.  Such a high neutrino luminosity is typical of core
collapse supernovae and protoneutron stars and can be therefore
detectable by the present neutrino telescopes for events occurring in
our galaxy. Notice that, differently from a supernova, the
process of conversion of a neutron star into a quark star would be a
powerful neutrino source but lacking an optical counterpart.
In general, a direct detection of neutrinos from protoneutron stars
would be an excellent opportunity to test 
the existence of quark matter in compact stars \cite{Pons:2001ar,Pagliara:2009dg,Pagliara:2010na} 

A final comment concerning the composition of the two stars is needed:
while the initial configuration, the neutron star, is rich of leptons (the electron 
fraction is of order $0.1$ at the center of the star), the final strange quark star
is almost lepton free. In the presently available simulations indeed 
no conservation of lepton number is imposed during the conversion. We expect
that, if included, the conservation of lepton number would lead to non-beta stable
strange quark matter. The cooling would be then accompanied by deleptonization
and the neutrino signal could be actually longer and more powerful.

\section{Conclusions}

A part from the neutrino signal itself, the process here considered
represents a new astrophysical energy source which could have a connection
with some of the most powerful explosions of the universe i.e. gamma-ray-bursts.
In particular, in some of these events the light curve shows some
non-trivial temporal structures, the so-called quiescent times.
These correspond to periods of tens of seconds during which 
it is likely that the inner engine of the gamma-ray-burst is not active at all \cite{Drago:2005rc}.
After the quiescent time the inner engine emits a new burst which has almost 
the same temporal and spectral structure of the first emission episode.
In a spectacular event detected by the Swift satellite, the GRB 110709B
\cite{Zhang:2011vk}, the quiescent time lasts about 10 min, 
which is probably a too long time of inactivity for a collapsar like inner engine.
It is now widely accepted that at least some gamma-ray-bursts 
are produced by highly magnetised protoneutron stars \cite{Metzger:2010pp}.
We speculate that a possible energy source for the second emission episode
seen in some light curves is the conversion of nucleonic matter to strange quark matter.
A detailed calculation showing that it is indeed possible to
emit part of energy released in the conversion in gamma-rays is clearly needed.


\end{document}